\documentclass[sigconf, nonacm]{acmart}

\usepackage{amsmath,amsfonts}
\usepackage{algorithmic}
\usepackage{graphicx}
\usepackage{textcomp}
\usepackage{xcolor}
\usepackage{booktabs} 
\usepackage{multirow}
\usepackage{xspace}
\usepackage{balance}
\usepackage{url}
\usepackage{subcaption}
\usepackage[font=footnotesize]{subcaption}
\usepackage[ruled,linesnumbered]{algorithm2e}
\usepackage[capitalize,noabbrev]{cleveref}
\usepackage[normalem]{ulem}

\newcommand{\name}{CEMENT\xspace}

\newcommand{\highlight}[1]{{\color{black}#1}}
\newcommand{\dfj}{Defects4J\xspace}

\AtBeginDocument{%
  \providecommand\BibTeX{{%
    \normalfont B\kern-0.5em{\scshape i\kern-0.25em b}\kern-0.8em\TeX}}}

\begin{document}

\title[Using Evolutionary Coupling to Establish Relevance Links Between Tests and Code Units. A case study on fault localization.]{Using Evolutionary Coupling to Establish Relevance Links Between Tests and Code Units. A case study on fault localization.}

\author{Jeongju Sohn}
\email{jeongju.sohn@uni.lu}
\affiliation{%
  \institution{Interdisciplinary Centre for Security, Reliability and Trust, University of Luxembourg}
  \country{Luxembourg}
}

\author{Mike Papadakis}
\email{michail.papadakis@uni.lu}
\affiliation{%
  \institution{Interdisciplinary Centre for Security, Reliability and Trust, University of Luxembourg}
  \country{Luxembourg}
}

\begin{abstract}
Many software engineering techniques, such as fault localization, operate based on relevance relationships between tests and code. These relationships are often inferred through the use of dynamic test execution information (test execution traces) that approximate the link between relevant code units and asserted, by the tests, program behaviour. Unfortunately, in practice dynamic information is not always available due to the overheads introduced by the instrumentation or the nature of the production environments. 
To deal with this issue, we propose \name, a static technique that automatically infers such test and code relationships given the projects' evolution. 
The key idea is that developers make relevant changes on test and code units at the same period of time,  i.e., co-evolution of tests and code units reflects a probable link between them.  We evaluate \name on 15 open source projects and show that it indeed captures relevant links. Additionally, we perform a fault localization case study where we compare \name with an existing Information Retrieval-based Fault Localization (IRFL) technique and show that it achieves comparable performance. A further analysis of our results reveals a small overlap between the faults successfully localized by the two approaches suggesting complementarity. In  particular, out of the 39 successfully localized  faults, two are common while \name  and IRFL localize 16 and 21. These results demonstrate that test and code evolutionary coupling can effectively support test and debugging activities. 

\end{abstract}




\maketitle
\pagestyle{empty}

\section{Introduction} 
\label{sec:introduction}

Software Engineering textbooks note that the majority of the total effort putted in a project during its life-cycle is during its maintenance phase~\cite{Tassey:2002rt}.
Consequently, many researchers are focusing on automating software maintenance related activities such as automated testing, test data generation and automated debugging~\cite{Wong:2016aa,Kamei2016:saner,Gazzola2019:apr,Fraser:2013:evosuite,Zeng:2016:fse,Harman:2018:scam}. These studies often rely on dynamic execution information, in order to identify links between tests and code. For instance, the studies on test selection often rely on test execution traces, to select tests that are likely to exercise the recently committed changes. 
Fault localization studies also adopt dynamic information to identify code locations that triggered test failures by narrowing down the code affected by a test failure~\cite{Wong:2016aa}. 

While precise, dynamic test execution information such test coverage is not always available, mainly due to the difficulty and the cost of data collection~\cite{Sohn2021ea,Memon:2017:seip,Zhang:2018:icse,Menta:2021:fse:industry,Bertolino:2020:icse}. 
For instance, test coverage, one of the most frequently used dynamic code analysis information, requires instrumentation, which may or may not be possible to perform in given environments. Even when the employed environment supports test coverage collection, this functionality is often turned-off, as it introduces significant overheads in order to log the related information~\cite{Sohn2021ea,Memon:2017:seip,Bertolino:2020:icse}.  Furthermore, when developers are under a fast-release cycle, they are unlikely to have enough time for data collection at the first place~\cite{Memon:2017:seip,Zhang:2018:icse,Elbaum2014lk}. 

To overcome the absence of dynamic coverage information, researchers have proposed static approaches that aim at guessing (predicting) the relations between tests and code~\cite{Saha:2013rm,Wen:2016:locus,Rompaey:2009:ecsmr,Cadar:2013:comm_acm,Blackshear:2018:RacerD}.
While encouraging, the working assumptions of these approaches are not often met. For instance, Information Retrieval based Fault Localization approaches \cite{Saha:2013rm,Wen:2016:locus,Koyuncu:2019:dc} assume the existence of bug reports, Focal method identification techniques \cite{Mohammad:2015:scam} assume particular test patterns \cite{WU:2020:jss} and similar naming convention that can be captured by string-matching \cite{White:icse:2020,Rompaey:2009:ecsmr,Zhang:2015:testName:ase}, assumptions that often are not met. 

We fill this gap by proposing \highlight{\name (CoEvolution between MEthod aNd Test)}, a static technique that automatically infers test and code relationships given the projects' historical  evolution. Thus, instead of relying on dynamic or static code analysis, \name it relies on how software has evolved. Precisely, \name establishes links, called Evolutionary Couplings  \cite{ZimmermannWDZ04}, between tests and code units by checking the tests and code that have co-evolved throughout the development. 

The key idea is that changes, on both tests and code, made around the same time imply a probable relevance coupling between them. Consider for example bug fixing cases, these are often followed or preceded by additions/modifications of tests (to reproduce and validate the repair action). Similarly, functionality additions are frequently followed by test additions. Therefore, the co-evolution analysis of tests and code units can capture a probable relevance coupling between them.


While there are many aspects of software evolution that could potentially  be exploited, i.e, commit time, developer or the context of the evolution, we stay simple by focusing only on whether tests and code have been altered in similar periods of time (e.g., within few commits to each other). 
\name differs from existing techniques as it is independent of the dynamic execution traces and the semantics of the source code. We posit that these differences allow test and code evolutionary couplings to complement existing techniques as they are capturing largely underexplored dependencies. Furthermore, we believe that co-evolution of tests and code units (i.e., methods) reveals important and hard to capture, by other techniques, couplings.

We empirically evaluate \name's ability to infer links between tests and code by investigating its ability to select relevant tests (for given code methods) and to perform fault localization, i.e., successfully localize faulty methods given failed and passing test cases. 
\highlight{We thus, inject some faults/mutants, in essence applying mutation testing \cite{PapadakisK00TH19}, on a set of selected methods and check the ability of \name to select tests that detect (kill) them.}  
Then we design a fault localization case study, where we investigate whether faulty methods have stronger links with failing tests than the passing tests. This means that we form a novel fault localization method, on top of \name, and compare its performance with that of an existing Information Retrieval-based Fault Localization technique~\cite{Koyuncu:2019:fse} that performs particularly well on the set of the projects that we study.  

Our results, conducted on the 15 open source projects of Defects4J v.2.0.0~\cite{Just:2014aa}, show that \name can infer relevant links between tests and code methods that can support software maintenance activities (e.g., fault localization) given the past co-evolution of the software.

In summary, the technical contributions of this paper are: 

\begin{itemize}
  \item The introduction of evolutionary coupling between tests and code, a novel type of coupling established based on how tests and code have co-evolved. Additionally, since the coupling between tests and code can be captured in a static way \name  offers advantages when dynamic information is not available.    

  \item Empirical evidence that \name can establish evolutionary couplings between tests and code. 
  Results from a fault localization case study show that \name can be useful in fault localization. 

  \item  Empirical evidence that evolutionary coupling between tests and code improves as software becomes more mature. The comparison between the \name's fault localization results for the projects with different levels of software maturity shows that \name becomes more effective when the project under inspection has actively evolved (i.e., changed).



  \item Empirical evidence that fault localization using \name complements state-of-the-art static fault localization techniques. Our results shows that \name can localize faults for which an existing fault localization technique has failed, implying the relevant links \highlight{captured by \name can complement this technique.}

\end{itemize}


\section{Evolutionary Coupling between Tests and Code} 
\label{sec:name_co_changes_of_test_and_code}

The key idea underlying our approach is that \highlight{developers make focused changes to their projects. Instead of making multiple irrelevant actions, they focus on one action at the time~\cite{Harman:2018:scam}. Even if the changes are not serving one purpose they are closely related since they are the result of the developer focus/attention.} 
This means that the changes committed around the same period are usually relevant to each other. For instance, when developers repair faults in code, they often introduce or modify a test to evaluate the repaired part. Similarly, when they implement new functionality or update existing ones, they probably introduce or update the tests related to this functionality, at the same time or shortly after. Even when some changes are irrelevant these should be eliminated through the number of evolutions/changes as it is unlikely to have  the same irrelevant changes repeatedly. Based on these, we formulate the following hypothesis, which forms the main idea of our work:\\ 

\begin{center}
\noindent\fbox{
  \parbox{0.44\textwidth}{
    \textbf{Hypothesis}: Changes in code units that are followed by the changes in tests (and vice versa) imply a relevance relationship (coupling) between them. Similarly, changes in code units not followed by changes in tests imply an absence of relevance. 
  }
}\newline
\end{center} 

We use the established term evolutionary coupling \cite{ZimmermannWDZ04} to name the coupling between tests and code established from the above hypothesis, i.e., guided by co-changes or, in other words, the co-evolution of tests and code. 
\highlight{This paper aims to investigate whether this evolutionary coupling between tests and code can be useful in software testing and debugging activities, more specifically, whether they can be used to fault localization.} 
%
Hence, we propose \name, a static approach that automatically infers links between tests and code units relying on evolutionary coupling. We work with methods since they form discrete and localizable units, typically targeted by automated techniques such as fault localization. 

\subsection{Tests and Code Evolutionary Coupling} 
\label{sub:evolutionary_couplings_between_tests_and_code}

Evolutionary coupling between tests and code exists when tests and code have co-evolved throughout the software development: it is independent of specific changes and simply focuses only on whether the changes in tests and code occurred in similar time periods. As a result, this coupling does not require any dynamic information to be established, hence being easier to exploit in some cases.
In addition, by relying on the timing when each test and method change was made, \highlight{the relation captured by the evolutionary coupling is independent of any software testing and debugging tasks, but reflects important links since they reflect the developer's intent as developers tend to do them together.} 
Another distinct characteristic of evolutionary coupling is that it is inferred from the projects' evolution and thus, it evolves along with the target software. The more changes performed to the software, or the more mature the software is, the better. This feature can be very helpful in development cases following the continuous integration development model, as they tend to evolve both tests and code at the same time and include finer-grained commits. Additionally, such evolutionary couplings can provide developers up-to-date guidance on which code is linked to which tests and vice versa easing comprehension.

\subsection{\name}
\label{sub:approach}

For a given set of methods and tests, \name identifies evolutionary couplings, between each method and test pair, by computing the average time interval between their past changes. Before going into the details of \name, we detail and show the distinct nature of tests and method coupling that \name aims for, which we call \textit{the coupling asymmetry}.

\subsubsection{Asymmetry in the test and method coupling}

In an ideal case where each test and method assesses and implements a unique functionality, the coupling between them is symmetric: a method is associated with a test at a degree equal to x\%, which is also equal to the degree the test associates to the method. However, in practice, we frequently observe cases where a single test exercises, directly or indirectly, (simply relates) to multiple functionalities. We also observe methods implementing multiple functionalities. For these cases, the coupling between methods and tests becomes asymmetric, \highlight{i.e., a test may relate with a method at a different degree than the method with that test.} 
The degree of the association reflects the strength of the relation between entire tests and code methods. For example, let us assume that method $m$ implements a functionality $f$ that is a part of a more extensive functionality $F$; test $t$ examines the functionality $F$, indirectly evaluating the functionality $f$. 

For method $m$, test $t$ is the most related one, as it evaluates its main functionality, $f$. However, test $t$ has a stronger degree of coupling to another method $m'$ that carries out the entire functionality $F$, making the coupling between method $m$ and test $t$ asymmetric. Still, test $t$ is the one to run when we need to inspect method $m$. \name takes into account  this asymmetry and computes a coupling degree separately for a method and for a test. It then aggregates these values from both test and method sides to estimate their final coupling degree. 

\subsubsection{Computing the distance between the test and method}
\name measures the degree of evolutionary coupling between the test and method as the time interval between their past changes. For the time interval between two changes, \name counts the number of commits between them. Hereafter, we will refer to this time interval as the distance. There are two additional points that we need to consider when calculating this distance. First, tests and methods are likely to be altered more than once, especially when the software under inspection has evolved actively. Secondly, the obtained distance is inherently asymmetric since it quantifies the degree of asymmetric coupling.

\cref{alg:pseudo_closest_distance} presents the pseudo-code of computing the distance from target $t$ to $tc$. Both $t$ and $tc$ can be either a method or a test. We denote as $t$ a test and $tc$ a method. 
Tests and methods are likely to be modified more than once from their introduction. Therefore, we first collect a list of commits that changed test $t$ and method $tc$ (Line 1 and 2). Here, we are interested in inspecting whether the test and method have been changed around the same time. Thus, we consider only the distance to the \textit{nearest} method change for each test change rather than taking all the past changes of the method into account (Line 3 to 7). If the test and method are related to each other, their changes can trigger the changes or be triggered by the changes of the other party. Thus, we use the absolute distance while we search for the nearest method change. We then aggregate these distances computed for each past test change by taking the average (Line 8). By using the average, we can reduce the risk of being affected by the outlier case where the method and test were accidentally modified at the same time period. 

The distance we define is asymmetric. For example, let us assume that method $M$ was altered at the commits $c_0$ and $c_3$ and test $T$ at the commits $c_1$, $c_3$, and $c_4$. For the method changes at $c_0$ and at $c_3$, the distance to the nearest test change is 1 and 0, respectively; the final distance of method $M$ to test $T$ is thereby $\frac{1+0}{2} = \frac{1}{2}$. However, for test $T$, the distance to method $M$ is $\frac{2}{3}$ and not $\frac{1}{2}$, as the shortest distance for individual test changes at $c_1$, $c_3$, and $c_4$ is 1 ($|-1|$), 0, and 1, resulting in the average distance of $\frac{2}{3}$. This asymmetry of the distance has occurred for two reasons. First, we consider only the nearest change, and while doing that, we search both back and forth, allowing each change to select any change as the nearest one regardless of whether the other party also considers it as the nearest. More importantly, the coupling between methods and tests is asymmetric. Since the distance is a way to quantify the degree of the coupling between the test and method, it naturally becomes asymmetric if the coupling is asymmetric; this is also why we did not define the distance to work in both ways. \name considers this asymmetry and calculates distances from both directions, i.e., from a test to a method and the opposite, when it estimates the coupling degree, i.e., the distance, between tests and methods.

\begin{algorithm}[t]
\SetAlgoLined
\SetKwInOut{Input}{input}\SetKwInOut{Output}{output}
\Input{
a target, \textbf{\textit{t}}, a comparison target, \textbf{\textit{t$_{c}$}}, a distance aggregation method, \textbf{\textit{M$_{avg}$}} 
}
\Output{
the distance of a target \textbf{\textit{$t$}} to the nearest changes in the comparison target \textbf{\textit{$t_{c}$}}
}

$Revsions_{t} \leftarrow$ ChangeTarget$(t)$%

$Revsions_{t_c} \leftarrow$ ChangeTarget$(t_c)$%

$Dists \leftarrow []$%

\For {$rev$ in $Revsions_{t}$}{
  $d \leftarrow min(\{$distance$(rev, rev') | rev' \in Revsions_{t_c}\})$ %

  add $d$ to $Dists$
}

dist$_{t, t_c} \leftarrow M_{avg}(Dists)$%

\Return dist$_{t, t_c}$%

\caption{DistanceToNearest}\label{alg:pseudo_closest_distance}
\end{algorithm}

\subsubsection{Establishing Links between Tests and Methods} 

\cref{alg:pseudo_ctcc} describes how \name infers \highlight{relevant} links between tests and methods from the distances calculated from \cref{alg:pseudo_closest_distance}. 
The final output of \name is a list of methods/tests ($C$) sorted in descending order of their degree of coupling to the test/method under inspection ($T$): the higher the rank of a method/test is, the stronger its link to the target test/method.
For this, \name first calculates the distance of target $T$ to each candidate in the list $C$ (Line 1). 
After computing the distance to the target for each candidate in $C$, we select the top $N$ that are closest to $T$ (Line 2). \name then calculates the distance to $T$ for each candidate in $C_N$ and multiplies newly obtained distances with their matching distances from the target (Line 4 to 6). 
\name then sorts the candidates in descending order using these updated distances; for those failed to be in the top $N$, we use their distance values in $D_{T \rightarrow C}$ (Line 3). 
This additional update for the top $N$ is to handle the asymmetric nature of the distance. We set $N$ to 100, which we obtained empirically. 
\highlight{To summarize, \name considers a method and a test to be relevant (likely coupled) if and only if both of them are considered to be close enough. This condition helps avoiding coincidental cases}\footnote{There is no need of having both methods and tests considering each other as the most likely-to-be relevant one. Being relatively close to each other, compared to the rest of tests/methods is sufficient to establish a link between them.}.

\begin{algorithm}[t]
\SetAlgoLined
\SetKwInOut{Input}{input}\SetKwInOut{Output}{output}
\Input{
a target, \textbf{\textit{T}}, a list of likely-relevant candidates, \textbf{\textit{C}}, a distance aggregation method, \textbf{\textit{M$_{avg}$}}, the number of top candidates, \textbf{\textit{N}}
}
\Output{
a list of candidates \textbf{\textit{C}} sorted in descending order of the distance to the target \textbf{\textit{T}}
}

$D_{T \rightarrow C} \leftarrow$ DistanceToNearest$(T,C, M_{avg})$ %

$C_{N} \leftarrow$ SelectTopN$(N, D_{T \rightarrow C}, C)$%

$D_{T \leftrightarrow C} \leftarrow$ Initialize$(D_{T \rightarrow C}, T, C)$%

\For {$c$ in $C_{N}$}{
  $d_{c \rightarrow T} \leftarrow $ DistanceToNearest$(c,T, M_{avg})$
  $D_{T \leftrightarrow C}[c] \leftarrow d_{c \rightarrow T} \cdot D_{T \leftrightarrow C}[c]$  
}%

$C_{ranked} \leftarrow$ Rank$(D_{T \leftrightarrow C}, C)$

\Return $C_{ranked}$

\caption{\name}\label{alg:pseudo_ctcc}
\end{algorithm}


\section{Experimental Settings} 
\label{sec:experimental_setting}


\subsection{Research Questions} 
\label{sub:research_questions}

To evaluate \name we start our investigation by checking its ability to mine true links between code units and tests. Thus, we ask:

\begin{description}
\item[RQ1] \textbf{Capability:} can \name establish static links between tests and methods based on their co-evolution?
\label{subsub:rq1}
\end{description}

To answer this question, we need an oracle that decides on the link between tests and methods. This type of oracle (i.e., general associations between tests and methods), however, is hard to get and there is no guarantee that it can be accurately approximated \cite{White:icse:2020}. To set such an oracle we use mutation testing \cite{PapadakisK00TH19} applied at specifically selected methods and check whether tests selected by \name can indeed kill the mutants of the selected methods and contrast them with randomly selected tests. 
The underlying assumption here is that tests related to the methods should kill the mutants that reside on these methods, at least kill more mutants, than tests that are not related. We thus, expect the resulting test-and-killed relations between tests and methods to overlap with the links inferred by \name.

Specifically, to answer this RQ, we select $N$ tests that are the most likely to kill mutants in the methods we consider by picking the top $N$ tests with the strongest coupling to these methods. 
\highlight{As \name establishes a relationship between a test and a method by default, we take additional steps to obtain a relationship between a test and multiple methods. To be specific, we first repeat \name for each method using all tests, obtaining multiple rankings for each test; we then take either the highest (i.e., the best) or the average as the final ranking for each test.}  
These two are notated as \name$_{t \rightarrow mm}^{best}$ and \name$_{t \rightarrow mm}^{avg}$; $t$ and $mm$ denote a test and multiple methods, respectively. 
We deem \name capable of inferring the links between tests and methods from their co-evolution if the tests ranked within the top $N$ by \name can kill more mutants than randomly selected $N$ tests.

After investigating the existence of links between tests and methods, we turn our attention to a more concrete task in order to investigate whether these links offer actionable information. Therefore, we ask:

\begin{description}
\item[RQ2] \textbf{Applicability:} can we use the links between tests and methods generated by \name in software debugging?
\label{subsub:rq2}
\end{description}

To answer this question, we conduct a case study of \name in fault localization.
We select fault localization among different software debugging activities since we can directly relate the resulting test and method links to fault localization by assuming the methods strongly coupled to a failing test as suspicious ones. 
We also compare \name with an existing Information Retrieval-based Fault Localization (IRFL) technique adopted in a recent program repair technique, iFixR~\cite{Koyuncu:2019:fse}. We choose this IRFL technique as our baseline because it combines various IR-based features of faults, summarising existing IRFL techniques.\footnote{iFixR performs statement-level fault localization by localizing faults at the file-level first using D\&C that leverages various IR features~\cite{Koyuncu:2019:dc}} In addition, the IRFL is a static approach, thereby allowing us to inspect further how \name performs compared to existing static techniques. 

After investigating the applicability of \name on fault localization we check whether its performance is dependent on the project maturity. Hence, we ask:

\begin{description}
\item[RQ3] \textbf{Impact of Software Maturity:} how does the software maturity affect the effectiveness of \name?
\label{subsub:rq3}
\end{description}

\name assumes tests and methods to be relevant if they have co-evolved throughout the development. Thus, for \name to be useful, the program under inspection should be mature enough to have a sufficient amount of previous changes for \name to process. 
Hence, to answer RQ3, we divide our dataset into two levels of software maturity (i.e., Applicable and Confident) based on the number of changes made on individual tests and methods. 

\begin{itemize}
  \item Applicable: tests and methods have changed at least once
  \item Confident: tests and methods have changed equal to or more than the average number of times individual tests and methods have changed 
\end{itemize}%

\highlight{We extend our fault localization case study to additionally have these two different levels of software maturity for each project and investigate how the performance of \name varies depending on the software maturity. 
We achieve this by using these two software maturity levels as the filtering criteria: \textit{Applicable} is to inspect the changes in \name's performance after excluding the tests and methods it cannot handle inherently, and \textit{Confident} is to simulate how \name performs when the projects become more mature.}  


\subsection{Subject} 
\label{sub:subject}

%

We evaluate \name using \dfj v.2.0.0~\cite{Just:2014aa}, a repository that contains real-world faults of 17 open source projects. We select \dfj mainly for two reasons. First, \dfj provides an inner command to run mutation testing on 17 projects it investigated. 
This inner command handles from compiling a project to running a given test suite on the injected mutants, saving us a lot of time and effort that would spend on preparing an environment for mutation testing. 
\highlight{Second, \dfj provides 835 real-world faults and an infrastructure to replicate them easily, making it an optimal target for our fault localization case study.
\cref{tab:subjects} presents the overall information about the projects that we studied: out of these 17 projects, we use 14 to answer RQ1 and 15 to answer RQ2 and RQ3.}

\subsubsection{Subjects for Mutation Testing}
\highlight{Among 17 projects in \dfj, we fail to run mutation testing on Commons Collections and Mockito using its inner mutation testing command. We simply exclude these two from our targets for mutating testing, as our goal is not about running mutation testing on every project in \dfj.  In this study, we use Git to collect the projects' past changes, as 16 out of 17 projects in \dfj employ Git to manage the changes between versions; JfreeChart is the only one that uses SVN instead.\footnote{Currently, JfreeChart has successfully migrated from SVN to Git. However, \dfj stills refers to the older version of JfreeChart and generates a new git branch that contains only four commits made by \dfj developers.} 
We thus, excluded JfreeChart, leaving 14 projects for our analysis.}

\subsubsection{Subjects for Fault Localization}

For our case study on fault localization, we exclude Commons Collections and JfreeChart. 
We exclude JfreeChart for the same reason above and Commons Collections for a different reason. 
\highlight{As explained in RQ2, we assume faulty methods to be more strongly related to failing tests than non-faulty ones. For \name to establish these relevant links between failing tests and methods, failing tests should exist before the failure. However, for many of the faults in \dfj, the failing tests marked by \dfj were introduced the first time to the project when developers submitted fix patches, making them impossible for \name to handle. \footnote{\dfj isolates each test failure and provides a separate commit that contains only a single fault and fails only with the tests related this fault. \dfj achieves this by reversing the bug-fixes and labeling the tests related to the fixes as the failing tests.} Thus, we further filter out them, excluding Commons Collections entirely since there were no remaining faults after this exclusion. Combined, we are left with 214 faults out of 835 faults.}

We use the IRFL in iFixR as the baseline for our fault localization case study. This IRFL technique also failed in JfreeChart for a similar reason to ours.
Among 214 faults, the IRFL failed to localize 39 faults due to missing bug reports or bug reports related to more than one fault and 33 faults by failing to compute suspiciousness scores for faulty statements. Thus, we further exclude 72 faults and use the remaining 142 faults for the baseline comparison in RQ2. \cref{tab:subjects} presents the number of faults remaining after these exclusions for each project. 
RQ3 is about the impact of software maturity on the performance of \name. We use all 214 faults to answer RQ3 since we do not need to compare \name with the IRFL here.

\begin{table}[ht!]
\centering
\caption{Test subjects. 
Mockito was excluded from RQ1. For RQ2 and 3, all 15 projects were used. The value in parentheses is the number of faults after removing those the IRFL cannot handle. 
\label{tab:subjects}}
\vspace{-0.8em}
\scalebox{0.78}{
\begin{tabular}{lr|rrrr}
\toprule
 & \# of & \# of methods & \# of tests & \multicolumn{2}{c}{\# of previous commits} \\
Project & faults & & & min -- max & average \\
\midrule 

Lang & 23 (21) & 1959.3 & 2201.6 & 3695 -- 4769 & 4160.9 \\
Math & 32 (24) & 4089.7 & 3288.1 & 3929 -- 10074 & 7378.2 \\
Time & 3 (2) & 3782.7 & 5112.7 & 8849 -- 8988 & 8895.3 \\
Closure & 60 (27) & 9857.9 & 8435.6 & 10332 -- 23795 & 18293.5 \\
Mockito & 8 (1) & 3118.4 & 2187.0 & 4296 -- 5808 & 5305.4 \\
Cli & 12 (7) & 216.9 & 228.2 & 324 -- 906 & 445.1 \\
Codec & 6 (4) & 408.0 & 537.7 & 587 -- 1301 & 945.7 \\
Compress (11) & 12 & 1202.0 & 675.6 & 569 -- 3077 & 1877.6 \\
Csv & 4 (3) & 109.5 & 205.2 & 147 -- 470 & 314.8 \\
Gson & 2 (2) & 805.0 & 1353.5 & 2157 -- 2160 & 2158.5 \\
JacksonCore & 6 (5) & 1718.0 & 722.3 & 1621 -- 3292 & 2440.3 \\
JacksonDatabind & 19 (15) & 5427.3 & 3358.3 & 7187 -- 10330 & 8786.3 \\
JacksonXml & 1 (0) & 388.0 & 298.0 & 686 -- 686 & 686.0 \\
Jsoup & 21 (18) & 985.0 & 446.0 & 639 -- 2134 & 1431.0 \\
JxPath & 5 (2) & 1518.2 & 475.4 & 1947 -- 2045 & 1993.6 \\

\bottomrule
\end{tabular}}
\vspace{-1.0em}
\end{table}  


\subsection{Past Change Collection} 
\label{sub:past_change_collection}

\name determines whether a given method and test are relevant using the time intervals between their past changes. Thus, to run \name, we first need to find when each method and test was changed. We employ Git v.2.32.0~\cite{chacon2014Git} for this because it is a widely used version control system and 15 projects we studied also utilize Git to maintain their versions. 
\highlight{For the experiments, we use \dfj, which works with the project's detached HEADs it created instead of the main branch of the project. Thus, to avoid confusion, such as the mismatch between commit hashes, we gather the past changes of methods and tests from these detached HEADs.}
This paper aims to demonstrate that CEMENT can infer the relevant links between tests and methods based on their history of co-evolution. Hence, rather than trying to achieve the best performance by controlling the time window for past change collection, we simply collect all prior changes of methods and tests. For each past change, we record the hash of the commit that introduced the change. 


\subsection{Mutation Testing}\label{sub:mutation_testing}

As mentioned in \cref{sub:subject}, \dfj provides an inner mutation testing command relying on Major. 
We thus, work with the buggy versions because it becomes easier to relate the results we obtained with mutation testing with those of the  fault localization case study we perform. Moreover, working with fixed versions may lead to overestimating the performance of \name, as the changes that generated the fixed versions may contain information that directly reveals the link between certain tests and methods (e.g., failing tests and faulty methods), which is rare in practice. We select the most recent version among multiple buggy versions for each project to maximize the number of past changes that \name can exploit.

Mutation testing is computationally expensive~\cite{PapadakisK00TH19}. 
Thus, we reduce the scope of the mutation testing by mutating only the top ten frequently modified classes instead of the entire code. We formulate a test set that includes 20\% of the entire tests of a target project. This test set includes all the tests that are likely to execute the selected classes according to the test and code naming convention~\cite{White:icse:2020}. In case there is available space after the selection process, we randomly select the remaining number of tests. 

\subsection{Fault Localization} 
\label{sub:fault_localization}

We conduct a case study on fault localization to evaluate the applicability of \name in software testing and debugging. As for the baseline of this case study, we select an Information Retrieval-based Fault Localization (IRFL) technique adopted in a recent program repair technique called iFixR~\cite{Koyuncu:2019:fse}; we run iFixR on 17 projects in \dfj v.2.0.0. 
Since we work at the method granularity, we aggregate the statement-level fault localization results of iFixR to the method level, taking the highest statement-level suspiciousness score for each method. 
For \name, we rank the methods in descending order of their distances to failing tests: the higher its rank is, the more suspicious the method is. \dfj includes faults that result in multiple failing tests. In these cases, we take the highest method ranking among those computed with each failing test; similarly, for multiple faulty methods, we take the highest.

While \dfj provides real faults, the corresponding buggy versions delivered by \dfj have been tailored to contain only a single fault~\cite{Just:2014aa}. Consequently, if we run \name directly on these buggy versions, we may include change information that explicitly reveals the locations of faults. To prevent this, instead of working with the buggy versions given by \dfj, we execute \name on the original buggy commits that \dfj additionally provides. For iFixR, we follow its own configuration.  


\subsection{Evaluation Metrics} 
\label{sub:evaluation_metrics}

We evaluate the effectiveness of \name in establishing the links between tests and methods from their evolutionary coupling using mutation score. Mutation Score (MS) is defined as the ratio of mutants killed by selected tests to the total number of generated mutants, as below. 
%
\highlight{\[MS(T) = \frac{\text{\# of mutants killed by $T$}}{\text{\# of total mutants}}\text{, } R_{MS}(T) = \frac{MS(T)}{MS_{max} (= MS(T_{all}))}\]}
%
$T$ denotes a test set that contains $N_T$ tests. We define $N_T$ (i.e., the number of selected tests) differently for each project, \highlight{configuring this $N_T$ to be 10\% of the total number of tests in the mutation testing. 
We compare the $MS$ score of test set $T$ with the $MS$ score of running all the tests ($T_{all}$).} 
We refer to the latter one as the maximum ($MS_{max}$); it is the total number of killed mutants over the total number of generated mutants. 
Here, we want to evaluate the trade-off between the effort saved by running only $N_T$ test and the performance degradation caused by it. Thus, we divide the $MS$ score of test set $T$ by the maximum $MS$ score. 
\highlight{We notate this ratio as $R_{MS}$ and use it to evaluate the capability of \name.}


RQ2 and RQ3 are about the applicability of \name in fault localization. We evaluate the fault localization performance of both \name and the baseline IRFL using $acc@n$ and $wef$. 
These two metrics measure the absolute effort spent on localizing faults, following the guideline suggested by Parnin and Orso~\cite{Parnin:2011uq}. $acc@n$ counts the number of faults ranked within the top $n$ places; for $n$, we use 1, 3, 5, and 10. $wef$ or wasted effort is the number of non-faulty elements examined before inspecting the first faulty one. As $wef$ is computed per fault, we take the average and the median.


\subsection{Tie Breaking} 
\label{sub:tie_breaking}

\name does not guarantee to calculate a unique distance (i.e., coupling degree) for each method and test pair; in the worst case, it may compute the same degree of coupling for all the method and test pairs. To avoid overestimating the performance, we break these potential ties between the method and test links by assigning the lowest ranking that tied methods or tests can have to all those that are tied. For a similar reasoning, we assign the lowest possible ranking to the methods tied by having the same suspiciousness score while evaluating the fault localization performance. 


\subsection{Implementation \& Environment}

\name is implemented in Python version 3.9.9. All the experiments were run on the machine equipped with Intel Core i7 CPU and 32GB RAM. The replication package and all the results are publicly available from \url{https://doi.org/10.5281/zenodo.6366615}.


\section{Results}\label{sec:results}


\subsection{RQ1. Capability} 
\label{sub:rq1_capability}

\cref{tab:rq1} records the changes in mutation score when executing subsets of the accompanied test suites, composed of 10\% of the total number of tests, using $R_{MS}$ score. Overall, we obtained higher $R_{MS}$ scores in \name (i.e., \name$_{t \rightarrow mm}^{best}$ and \name$_{t \rightarrow mm}^{avg}$) than in the random test selection.  
For example, for JacksonCore, $R_{MS}$ improves from 0.19 to 0.60, increasing the number of killed mutants more than three times. Among 14 projects we investigated, \name successfully outperforms the random test selection baseline in 11 projects by taking the highest ranking for each test among those it obtained with the considered methods (\name$_{t \rightarrow mm}^{best}$). 
When using the average instead of the highest (\name$_{t \rightarrow mm}^{avg}$), the number of projects where \name is superior than the random selection decreases by two. However, \name$_{t \rightarrow mm}^{avg}$ still performs better than the random in nine out of 14 projects. 
In fact, \name consistently outperforms the random test selection either by \name$_{t \rightarrow mm}^{best}$ or \name$_{t \rightarrow mm}^{avg}$.

The maximum Mutation Score (MS) assumes running all tests participated in the mutation testing. 
Since we selected 10\% of these tests, all the $R_{MS}$ scores in \cref{tab:rq1} being greater than 0.1, even in the random, suggests that there exist large overlaps between the tests in terms of the mutants they killed. 
Nonetheless, CEMENT were able to select 10\% of the tests that killed much more mutants than the same number of randomly selected tests. For instance, in JacksonXml, we can kill 74\% of the total killed mutants by running tests selected by \name, whereas with the randomly selected tests, we can kill only 28\%. 
Based on these results of \name consistently outperforming the random, we posit that \name can establish relevant links between tests and methods given the project's evolution.  \\

\textbf{Answer to RQ1:}
Tests selected by \name consistently kill more mutants than those killed by randomly selected tests. These results suggest that \name successfully captures probable links between tests and methods that can identify tests related to mutated methods.

\begin{table*}[ht]
\vspace{0.5em}
\centering
\caption{
The changes in mutation scores when running only 10\% of the entire tests. 
Each cell contains \highlight{$R_{MS}$,} the ratio of the mutation score obtained by the approach (row) to the maximum mutation score. When \name (i.e., \name$_{t \rightarrow mm}^{best}$ and \name$_{t \rightarrow mm}^{avg}$) outperforms the random test selection (Random), the corresponding ratio ($R_{MS}$) is highlighted in bold. For Random, its $R_{MS}$ score is highlighted in bold if and only if it outperforms both \name$_{t \rightarrow mm}^{best}$ and \name$_{t \rightarrow mm}^{avg}$.
}\label{tab:rq1}
\vspace{0.5em}
\scalebox{0.8}{
\begin{tabular}{l|rrrrrrrrrrrrrr}
\toprule
&  &  &  &  &  &  &  &  &  & Jackson & Jackson & Jackson &  & \\
Approach & Lang & Math & Time & Closure & Cli & Codec & Compress & Csv & Gson & Core & Databind & Xml & Jsoup & JxPath\\

\midrule
Random & 0.22 & 0.31 & 0.47 & 0.43 & 0.28 & 0.36 & 0.30 & 0.49 & 0.20 & 0.19 & 0.15 & 0.28 & 0.49 & 0.29\\
\midrule
\name$_{t \rightarrow mm}^{best}$ & 0.22 & \textbf{0.40} & \textbf{0.58} & \textbf{0.49} & \textbf{0.33} & \textbf{0.72} & 0.18 & \textbf{0.67} & \textbf{0.38} & \textbf{0.60} & 0.16 & \textbf{0.74} & \textbf{0.52} & \textbf{0.45}\\
\name$_{t \rightarrow mm}^{avg}$  & \textbf{0.54} & \textbf{0.38} & 0.36 & 0.40 & \textbf{0.52} & \textbf{0.47} & \textbf{0.49} & \textbf{0.65} & 0.14 & 0.13 & \textbf{0.33} & \textbf{0.64} & 0.48 & \textbf{0.49}\\

\bottomrule
\end{tabular}}
\vspace{0.5em}
\end{table*}  


\subsection{RQ2. Applicability} 
\label{sub:rq2_applicability}

\cref{tab:rq2_perf} presents the results of our case study of \name on fault localization. Compared to the Information Retrieval-based Fault Localization (IRFL) technique used in a recent program repair method, iFixR, \name acquires comparable performance in terms of $acc@1$: \name places 18 faults at the top of the rankings, whereas the IRFL places 23 faults. For the sake of simplicity, we will call the IRFL technique used in iFixR \textit{the IRFL}, hereafter. While the IRFL surpasses \name consistently in localizing faults, it also requires more effort to build the FL model and collect data for training and evaluation: the IRFL collects 17 features (seven from bug reports and ten from source code files) for fault localization. 
Moreover, the IRFL assumes the existence of bug reports that in many cases is not available and often requires an additional data preprocessing step, which can be costly~\cite{Koyuncu:2019:fse}.
\highlight{In contrast, \name needs only the hashes of commits that changed the methods and tests.
Hence, concerning the cost spent up to the localization,}\footnote{\highlight{With \name, each fault localization task itself was done within seconds, and the past change collection, which covers over thousands of commits here, took on average within 5 minutes, without any optimization. This cost can be further reduced in practice, as we do not have to process all prior commits every time; the changes can be collected incrementally.}} we posit that \name shows comparable performance to the IRFL.

An important discrepancy in the evaluation of IRFL happens when bug reports explicitly specify the code elements that have the reported bugs. 
In these cases, we do not need to localize faults in the first place, as they have been already identified by the person reporting them. The IRFL localizes faults based on the similarity between bug reports and source code. Therefore, if a bug report already contains the identifier of a fault, we might overestimate the performance of the IRFL, especially when it directly exploits the identifiers of code elements in a bug report. Hence, we divide faults into two groups based on whether their identifiers are already in bug reports and examine whether the localization performance differs between these two groups. 
We treat a bug report to contain the identifier of a fault if it has both the class and the method name of the fault. 

The leftmost column of \cref{tab:rq2_perf} presents the localization results of the faults whose identifiers are in bug reports; out of 142 faults we examined, 54 already have their identifiers in bug reports. The middle column shows the localization results of the faults without their identifiers in bug reports; the rightmost column describes the combined results. Overall, the IRFL performs better when bug reports contain fault identifiers: the IRFL places 15 out of 54 faults at the top ($acc@1$) for the group of faults with their identifiers in bug reports, whereas it places only eight at the top for the other group without the identifiers, even though more faults belong to the latter group. We observe similar trends in $acc@3,5,10$. 

Compared to the IRFL, having fault identifiers in bug reports does not have the same effect on \name. While \name localizes more faults near the top for the group where bug reports have fault identifiers, the difference is smaller; 
for example, for the group without fault identifiers, the $acc@1$ decreases almost by half in the IRFL, whereas, in \name, it decreases only by two, from 10 to 8. Even this small decrease is from elsewhere, as \name does not leverage both source code and bug reports in the first place; we suspect that the observed decreases are coincidental and are attributed to the characteristics of faults in each group.\footnote{If a method evolves actively, this method is more likely to have failed in the past than those rarely changed. Subsequently, if the method frequently fails, a reporter may already know that this method is the trigger when it causes a failure, and thereby, includes its identifier in the bug report.}
Furthermore, \name becomes more comparable to the IRFL for the faults without their identifiers in the bug reports: compared to the IRFL, \name ranks the same number of faults at the top and within the top three (i.e., $acc@1$ and $acc@3$) and locates only one less fault within the top five (i.e., $acc@5$).
\name fails to compete with the IRFL in $wef$, although the difference between \name and the IRFL becomes smaller in the median compared to the average. 
Regarding the previous observation in $acc@n$, this result is likely from \name completely failing on some faults, assigning the lowest rank to them. 
Nevertheless, \name still achieves comparable performance in terms of localizing faults near the top, suggesting that \name can be useful in fault localization.\\ 

\textbf{Answer to RQ2:} \name achieves comparable performance to the recent Information Retrieval-based Fault Localization (IRFL) technique, especially for the cases where this IRFL technique performs less effective.

\begin{table*}[ht]
\centering
\caption{Comparison between the fault localization by \name and the IRFL in iFixR. \name becomes more comparable to the IRFL when focusing only on the faults whose identifiers are not already in bug reports (Without Faulty Methods). 
}\label{tab:rq2_perf}
\scalebox{0.68}{
\begin{tabular}{l|rrrrrr|rrrrrr|rrrrrr}

\toprule
& \multicolumn{6}{c|}{With Faulty Methods (\name / iFixR)} & \multicolumn{6}{c|}{Without Faulty Methods (\name / iFixR)} & \multicolumn{6}{c}{All (\name / iFixR)} \\
 & \multicolumn{4}{c}{acc} & \multicolumn{2}{c|}{wef} & \multicolumn{4}{c}{acc} & \multicolumn{2}{c|}{wef} & \multicolumn{4}{c}{acc} & \multicolumn{2}{c}{wef} \\

Proj. (w/wo/all) & @1 & @3 & @5 & @10 & mean & med & @1 & @3 & @5 & @10 & mean & med  & @1 & @3 & @5 & @10 & mean & med\\

\midrule

Lang (16/5/21) & 3/7 & 8/9 & 9/11 & 10/12 & 60.9/4.2 & 2.0/1 & 0/0 & 2/1 & 3/2 & 3/3 & 441.8/11.8 & 3.0/6 & 3/7 & 10/10 & 12/13 & 13/15 & 151.6/6.0 & 3.0/3 \\

Math (12/12/24) & 1/2 & 1/6 & 2/7 & 3/11 & 562.0/4.0 & 52.0/2 & 0/3 & 0/6 & 0/8 & 0/11 & 787.8/11.6 & 86.0/2 & 1/5 & 1/12 & 2/15 & 3/22 & 674.9/7.8 & 83.0/2 \\

Time (1/1/2) & 1/0 & 1/1 & 1/1 & 1/1 & 0.0/1.0 & 0.0/1 & 0/0 & 0/0 & 1/0 & 1/0 & 4.0/60.0 & 4.0/60 & 1/0 & 1/1 & 2/1 & 2/1 & 2.0/30.5 & 2.0/30 \\

Closure (1/26/27) & 0/0 & 0/0 & 0/0 & 0/0 & 4315.0/20.0 & 4315.0/20 & 2/1 & 3/2 & 4/2 & 6/4 & 2225.6/60.3 & 242.0/22 & 2/1 & 3/2 & 4/2 & 6/4 & 2303.0/58.8 & 301.0/22 \\

Mockito (0/1/1) & - / - & - / - & - / - & - / - &-/- & -/- & 0/0 & 1/0 & 1/0 & 1/0 & 1.0/10.0 & 1.0/10 & 0/0 & 1/0 & 1/0 & 1/0 & 1.0/10.0 & 1.0/10 \\

Cli (3/4/7) & 2/1 & 2/3 & 2/3 & 2/3 & 18.0/1.0 & 0.0/1 & 2/0 & 3/1 & 3/1 & 3/1 & 4.2/13.8 & 0.0/17 & 4/1 & 5/4 & 5/4 & 5/4 & 10.1/8.3 & 0.0/2 \\

Codec (3/1/4) & 0/2 & 0/3 & 0/3 & 1/3 & 98.7/0.7 & 96.0/0 & 0/0 & 0/0 & 0/0 & 0/0 & 212.0/16.0 & 212.0/16 & 0/2 & 0/3 & 0/3 & 1/3 & 127.0/4.5 & 146.0/1 \\

Compress (6/5/11) & 0/1 & 2/1 & 3/2 & 4/2 & 152.2/19.7 & 5.0/14 & 0/0 & 0/0 & 0/0 & 0/2 & 55.6/16.6 & 50.0/19 & 0/1 & 2/1 & 3/2 & 4/4 & 108.3/18.3 & 21.0/18 \\

Csv (0/3/3) & - / - & - / - & - / - & - / - &-/- & -/- & 0/2 & 1/3 & 1/3 & 1/3 & 22.3/0.3 & 18.0/0 & 0/2 & 1/3 & 1/3 & 1/3 & 22.3/0.3 & 18.0/0 \\

Gson (2/0/2) & 1/0 & 1/0 & 1/0 & 1/0 & 15.0/52.5 & 15.0/52 & - / - & - / - & - / - & - / - &-/- & -/- & 1/0 & 1/0 & 1/0 & 1/0 & 15.0/52.5 & 15.0/52 \\

JacksonCore (1/4/5) & 0/0 & 0/0 & 0/0 & 0/1 & 1022.0/9.0 & 1022.0/9 & 2/0 & 3/0 & 3/0 & 3/0 & 77.2/130.0 & 0.0/127 & 2/0 & 3/0 & 3/0 & 3/1 & 266.2/105.8 & 1.0/16 \\

JacksonDatabind (5/10/15) & 0/0 & 0/2 & 0/3 & 0/3 & 620.2/20.0 & 398.0/4 & 0/1 & 0/2 & 0/2 & 0/4 & 1819.5/38.4 & 1044.0/12 & 0/1 & 0/4 & 0/5 & 0/7 & 1419.7/32.3 & 540.0/11 \\

Jsoup (4/14/18) & 2/2 & 2/3 & 2/3 & 3/3 & 170.5/8.5 & 4.0/1 & 2/1 & 3/1 & 4/3 & 5/7 & 215.1/35.9 & 81.0/10 & 4/3 & 5/4 & 6/6 & 8/10 & 205.2/29.8 & 39.0/7 \\

JxPath (0/2/2) & - / - & - / - & - / - & - / - &-/- & -/- & 0/0 & 0/0 & 0/0 & 0/1 & 279.5/120.0 & 280.0/120 & 0/0 & 0/0 & 0/0 & 0/1 & 279.5/120.0 & 280.0/120 \\

\midrule
Total (54/88/142) & 10/15 & 17/28 & 20/33 & 25/39 & 335.8/9.4 & 14.0/2 & 8/8 & 16/16 & 20/21 & 23/36 & 1047.5/41.3 & 80.0/12 & 18/23 & 33/44 & 40/54 & 48/75 & 776.8/29.2 & 63.0/8 \\

\bottomrule
\end{tabular}}
\end{table*}

\subsection{RQ3. The Impact of Software Maturity} 
\label{sub:rq3_the_impact_of_software_maturity}


\cref{tab:rq3_fl} presents the fault localization results of \name for the complete set of 214 faults with two variations on software maturity: \textit{Applicable} and \textit{Confident}. 
Methods or tests being "Applicable" means they have changed at least once, and being "Confident" implies that they have been altered more frequently than the average.  
We apply these two maturity criteria to each buggy version of target projects, filtering out the methods and tests that failed to meet them. We did not differentiate faulty methods and failing tests from other methods and tests while applying these criteria. As a result, 14 and 129 faults out of 214 faults were excluded by \textit{Applicable} and \textit{Confident} criteria, respectively. 

The results in \cref{tab:rq3_fl} show that the performance of \name can be improved as the software becomes more mature. For instance, when we apply \textit{Applicable} criterion (\textit{Applicable*}), we observe small improvements in the percentage of faults localized near the top, that are around 1\% to 3\%. The improvement is more evident in $wef$ where the average decreases by half and the median by around 15. When we further filter out immature methods and tests using \textit{Confident} criterion (\textit{Confident*}), this improvement becomes more prominent: the percentage of localized faults further increases by 6\% at the top and by around 6 to 8\% within the top three, five and ten compared to the results of applying \textit{Applicable} criterion. In $wef$, the average and the median decrease by around one fourth when compared to those of the \textit{Applicable}.

We exclude the methods and tests that failed to meet our maturity criteria to allow \name running on software systems with different maturity levels. Since we use absolute metrics (i.e., $acc@n$ and $wef$) in the evaluation, we consider the possibility that the performance improvement that we observed may come from the decrease in the total number of methods and tests after the filtering. To verify that the improvement comes from the project maturity, we apply the maturity criteria \textit{only} on faulty methods and failing tests, excluding faults that failed to meet these criteria. We then rerun \name for the remaining faults \textit{without} any maturity filtering. This way, we focus on faults that appeared at the mature part of the code without risking to overestimate. As shown in \cref{tab:rq3_fl}, our results are almost the same as before, localizing one more or fewer faults near the top, confirming that the improvement we witnessed 
indeed relates to the software maturity. \\

\textbf{Answer to RQ3:} 
The performance of \name improves when we focus only on the mature part of the code (i.e., have more than the average number of past changes), implying \name can enhance along with software maturity.

\begin{table}[ht!]
\centering
\caption{Complete fault localization results of \name. $N$ is the total number of faults in each project. "*" means that \textit{Applicable}/\textit{Confident} filtering criterion was applied to the projects before extracting the evolutionary couplings, and without "*" indicates that the filtering criterion was used only to exclude faults. 
\label{tab:rq3_fl}}
\scalebox{0.70}{
\begin{tabular}{lr|rrrrrr}
\toprule
Project & $N$ & \multicolumn{4}{c}{acc@n} & \multicolumn{2}{c}{wef} \\
& & 1 & 3 & 5 & 10 & mean & median \\

\midrule
Lang & 23 & 3 (0.13) & 10 (0.43) & 12 (0.52) & 13 (0.57) & 145.3 & 4.0 \\
Math & 32 & 2 (0.06) & 3 (0.09) & 4 (0.12) & 7 (0.22) & 692.0 & 79.5 \\
Time & 3 & 1 (0.33) & 1 (0.33) & 2 (0.67) & 2 (0.67) & 867.3 & 4.0 \\
Closure & 60 & 7 (0.12) & 11 (0.18) & 12 (0.20) & 16 (0.27) & 2130.1 & 228.5 \\
Mockito & 8 & 2 (0.25) & 3 (0.38) & 3 (0.38) & 5 (0.62) & 58.1 & 7.0 \\
Cli & 12 & 5 (0.42) & 7 (0.58) & 8 (0.67) & 8 (0.67) & 10.4 & 1.0 \\
Codec & 6 & 0 (0.00) & 0 (0.00) & 1 (0.17) & 2 (0.33) & 97.7 & 85.0 \\
Compress & 12 & 0 (0.00) & 2 (0.17) & 3 (0.25) & 4 (0.33) & 101.3 & 23.0 \\
Csv & 4 & 0 (0.00) & 1 (0.25) & 1 (0.25) & 1 (0.25) & 33.2 & 32.5 \\
Gson & 2 & 1 (0.50) & 1 (0.50) & 1 (0.50) & 1 (0.50) & 15.0 & 15.0 \\
JacksonCore & 6 & 2 (0.33) & 3 (0.50) & 3 (0.50) & 3 (0.50) & 431.5 & 154.5 \\
JacksonDatabind & 19 & 0 (0.00) & 0 (0.00) & 0 (0.00) & 0 (0.00) & 1369.1 & 540.0 \\
JacksonXml & 1 & 0 (0.00) & 0 (0.00) & 1 (1.00) & 1 (1.00) & 3.0 & 3.0 \\
Jsoup & 21 & 4 (0.19) & 5 (0.24) & 6 (0.29) & 8 (0.38) & 221.6 & 64.0 \\
JxPath & 5 & 0 (0.00) & 0 (0.00) & 0 (0.00) & 0 (0.00) & 444.4 & 400.0 \\

\midrule
Total & 214 & 27 (0.13) & 47 (0.22) & 57 (0.27) & 71 (0.33) & 906.2 & 64.5 \\

Total (Applicable) & 198 & 27 (0.14) & 47 (0.24) & 57 (0.29) & 71 (0.36) & 432.2 & 50.5 \\
Total (Applicable*) & 198 & 27 (0.14) & 47 (0.24) & 56 (0.28) & 71 (0.36) & 432.2 & 50.5 \\

Total (Confident) & 85 & 16 \textbf{(0.19)} & 27 \textbf{(0.32)} & 32 \textbf{(0.38)} & 38 \textbf{(0.45)} & \textbf{180.5} & \textbf{16.0} \\
Total (Confident*) &  85 & 17 \textbf{(0.20)} & 27 \textbf{(0.32)} & 31 \textbf{(0.36)} & 37 \textbf{(0.44)} & \textbf{115.3} & \textbf{14.0} \\

\bottomrule
\end{tabular}}
\end{table}


\section{Discussion}\label{sec:discussion}

\subsection{The Impact of Evolutionary Couplings in Software Debugging} 
\label{sub:the_impact_of_evolutionary_couplings_in_software_debugging}

\name establishes relevance links between tests and code only when they have co-evolved. Thus, \name may complement existing software debugging and testing techniques that rely on dynamic or static code analysis. 
To check for this potential complementarity, we revisit the results of RQ2, but at this time, inspect which faults are localized by \name and by the baseline IRFL technique.

\cref{tab:rq2_diff_simple} reports the number of faults localized by CCCT, the IRFL, and both. 
Overall, \name and the IRFL localize different faults near the top. 
For the groups of faults whose identifiers are already in bug reports, the faults localized at the top by both \name and the IRFL is only two; \name and the IRFL locate respectively eight and 13 different faults at the top. 
While there are more in common between the faults localized by \name and the IRFL within the top three, five and ten, many faults are still localized by only one of them. 
In cases where bug reports do not contain fault identifiers, fewer faults are localized by both techniques. At the same time, the number of faults localized only by \name now becomes similar to that of the IRFL; for example, both \name and the IRFL localize eight different faults at the top ($acc@1$). 
In total, \name localizes 16 faults for which IRFL has failed at the top;  this is around 41\% of total faults ranked at the top by either approach. Within the top ten, 22 out of 97 faults are localized exclusively by \name.
These results indicate that the relationships between tests and code captured by \name through the software co-evolution are different from those exploited in the current IR-based fault localization techniques.
Thus, we posit that \name has the potential to complement current software debugging and testing activities by establishing the relationships that have remained underexplored.

\begin{table*}[ht]
\centering
\caption{Comparison between faults localized by \name and the IRFL in iFixR. The value on the right (\textit{either}) is the total number of faults localized by either \name or the IRFL (i.e., union), whereas the values on the left denote the number of faults localized only by \name, only by the IRFL, and by both (i.e., intersection), respectively. When \name/the IRFL localizes faults on which the other failed, the corresponding cell is highlighted in bold text.
}\label{tab:rq2_diff_simple}
\scalebox{0.8}{
\begin{tabular}{lr|rrrrrrrr}
\toprule
& N & \multicolumn{2}{c}{acc@1} & \multicolumn{2}{c}{acc@3} & \multicolumn{2}{c}{acc@5} & \multicolumn{2}{c}{acc@10} \\
& & \name / iFixR / both & either & \name / iFixR / both & either & \name / iFixR / both & either & \name / iFixR / both & either \\
\midrule


With Faulty Methods & 54 & \textbf{8}/\textbf{13}/2 & 23 & \textbf{6}/\textbf{17}/11 & 34 & \textbf{8}/\textbf{21}/12 & 41 & \textbf{8}/\textbf{22}/17 & 47\\
Without Faulty Methods & 88 & \textbf{8}/\textbf{8}/0 & 16 & \textbf{14}/\textbf{14}/2 & 30 & \textbf{15}/\textbf{16}/5 & 36 & \textbf{14}/\textbf{27}/9 & 50\\

\midrule
All & 142 & \textbf{16}/\textbf{21}/2 & 39 & \textbf{20}/\textbf{31}/13 & 64 & \textbf{23}/\textbf{37}/17 & 77 & \textbf{22}/\textbf{49}/26 & 97\\

\bottomrule
\end{tabular}}
\end{table*}  


\subsection{Evolutionary Couplings and Traceability Links between Tests and Code} 
\label{sub:evolutionary_couplings_and_traceability_links_between_tests_and_code}

There are some studies specialized in modelling test-and-tested relationships among various relationships that tests and code can have~\cite{White:icse:2020,Qusef:2011:icsm,Rompaey:2009:ecsmr}. These relationships are called test-to-code traceability links and aim at capturing the intent of tests, i.e., identify the key functionality that is tested/asserted by a test, leaving out indirectly tested functionality. This means that the traceability links are abstract and not exact. Compared to test-to-code traceability links, the evolutionary couplings established by \name reflect a more relaxed relation. For example, let us suppose the given test and method belong to the same component and thereby have frequently changed around the same time by developers. In this case, even if the test does not directly test the method, \name will likely regard it as relevant to the method since they have co-evolved. 
To further examine how \name differs from or relates to exiting studies of test-to-code traceability links, we employ TCTRACER, a state-of-the-art approach that automatically establishes test-to-code traceability links~\cite{White:icse:2020}. We replicate the method-level traceability links prediction study in the TCTRACER paper and investigate whether \name can predict these links.

\cref{tab:tctracer} compares the performance of \name and TCTRACER in predicting test-to-code traceability links at the method level.\footnote{The results of TCTRACER provided by the authors contain only method and test pairs predicted to have a traceability link rather than the complete prediction results. Hence, we only investigate precision, recall, and F1 score and exclude AUC and MAP for this replication study.} 
Here, we assume that \name generates a stronger link for a method to the test that evaluates its functionality than to those that do not.
Thus, for each test, we rank methods in descending order of their strength of the link to it. We then simply regard the top five methods as having traceability links with the test. 
The performance of \name varies depending on the software maturity. Thus, we extend this study by applying the two software maturity criteria (i.e.,\textit{Applicable} and \textit{Confident}) in order to focus on the traceability links coming from the more mature part of the code. We achieve this by excluding the traceability links that failed to meet the maturity criteria from the evaluation. \cref{tab:tctracer} reports the number of test-to-code traceability links we initially have and the number of links after the filtering.\footnote{Because the current \name implementation handles only methods and not constructors for Java, we exclude two additional traceability links for each project}

TCTRACER generally outperforms CEMENT in predicting test-to-code traceability links. We believe that this due to the following three reasons. First, we simply take the top five methods for the prediction. Thus, even if \name ranks a related method at the top, we end up with four false positives, explaining the low performance, especially in precision. Secondly, because the oracle of test-to-code traceability links was formulated manually, it might be biased toward the methods and tests with similar names. This may give some advantages to TCTRACER, for it leverages the textual similarity between the names. Finally and most importantly, a test and a method can have an evolutionary coupling between them without being considered in the test-and-tested relationship as it could be an indirect link. Despite these, \name excels TCTRACER in Chart when inspecting only the links that met the \textit{Confident} criterion: out of the remaining four test-to-code traceability links, two of them are correctly predicted only by \name.

When we inspected these cases, we found that these two are when there are no common terms between test and method names and when a test calls multiple methods, especially after calling the related method. Since five out of eight techniques that TCTRACER combines compare test and method names in order to predict the traceability links, TCTRACER could be less effective if the test and method have entirely different names. TCTRACER employs techniques that exploit dynamic execution traces, such as Last Call Before Assertion, to complement this weakness. However, like the second case that we observed, if there are many methods between the test and the ground-truth method in an execution trace, TCTRACER becomes less successful in the prediction. 
\name leverages neither the dynamic execution traces nor the source code. As a result, it is inherently free from all the issues that may arise from using them; this allows \name to handle the cases for which TCTRACER has failed. Hence, we argue that \name can complement existing techniques of predicting test-to-code traceability links, especially when working with mature software systems.

\begin{table}[t]
\centering
\caption{
Comparison between \name and TCTRACER. 
The three values next to the project are the number of traceability links without any filtering, with \textit{Applicable} and \textit{Confident} filtering.
The left and right values are the results of \name and TCTRACER, respectively, and they are all in percentage.
}\label{tab:tctracer}
\vspace{-0.8em}
\scalebox{0.69}{
\begin{tabular}{l|rrr|rrr|rrr}
\toprule
 & \multicolumn{3}{c}{Lang (74/44/7)} & \multicolumn{3}{c}{IO (40/40/14)} & \multicolumn{3}{c}{Chart (35/25/4)}\\
Conf & Prec & Recall & F1 & Prec & Recall & F1 & Prec & Recall & F1 \\
\midrule 

All & 10 / 86 & 16 / 78 & 12 / 82 & 9 / 67 & 22 / 82 & 13 / 74 & 5 / 23 & 9 / 74 & 6 / 35\\
Applicable & 12 / 59 & 26 / 89 & 17 / 71 & 9 / 67 & 22 / 82 & 13 / 74 & 5 / 17 & 12 / 76 & 7 / 28\\
Confident & 8 / 12 & 22 / 89 & 12 / 21 & 6 / 29 & 14 / 100 & 9 / 45 & 21 / 2 & 75 / 50 & 33 / 4\\

\bottomrule
\end{tabular}}
\end{table}  





\section{Threats to Validity} 
\label{sec:threats_to_validity}

A primary threat to validity of our work is the absence of the oracle for the evolutionary couplings between tests and methods. To mitigate this threat, we used mutation testing as a substitute of this oracle. Since mutation testing has been often employed as a test oracle~\cite{PapadakisK00TH19}, we posit that it can also be useful for our case. In addition to the capability of \name in establishing the evolutionary couplings, this study also investigates the usefulness of the resulting couplings in software maintenance activities. For this, we select fault localization as our target for the case study, as it is one of the most actively studied areas in software maintenance~\cite{Wong:2016aa}. We select a recent Information Retrieval (IR) based fault localization technique as the baseline because it combines multiple existing IR-based techniques and, thereby, can summarize the current trend in fault localization to some extent.~\cite{Koyuncu:2019:dc,Koyuncu:2019:fse}.

The threats to external validity relate to whether our findings on the effectiveness of the evolutionary coupling can be generalized to other projects. We use 15 open source projects in \dfj, a benchmark of real-world faults, as our targets for evaluation. Still, additional studies on industrial projects may be needed to fully verify our hypothesis. 

Threats to construct validity relate to the evaluation metrics we use. To assess the capability of \name to select likely-related tests to given methods, we employ mutation score, a widely adopted metric in mutation testing~\cite{PapadakisK00TH19}. For the case study, we select three absolute evaluation metrics that have been frequently employed in fault localization~\cite{Parnin:2011uq,Wang:2015xe,Sohn2019jk,Lou:2021:fse}.


\section{Related Work} 
\label{sec:related_work}

Our definition of co-evolution depends on the past changes in tests and methods. 
Several studies have leveraged past changes in tests and methods. Defect prediction aims at predicting faults before executing them using code quality metrics that include past changes in the software~\cite{McIntosh:2018:tse,Rahman:2013:icse,Kamei:2013:tse}. However, these usages of past changes are often limited to enrich the description of individual code elements concerning a specific problem: e.g., methods that have changed frequently are more likely to contain faults than those that have not~\cite{Rahman:2013:icse}. In other words, they use the past changes to describe the characteristics of faults. 

Sohn and Yoo \cite{Sohn2019jk,Saha2019jh} considered the past changes of methods to improve fault localization performance; similarly to defect prediction, they used the past changes as another feature to describe faults. In automated program repair, Saha et al. used past software changes to further locate the code that is likely to undergo similar repairs~\cite{Saha2019jh}. Although the ways they used the past changes varies, all these studies utilize past software changes as additional data to enhance their approaches; they can validate their main idea without using the past changes. Furthermore, they inspect past changes per code element rather than investigating them together to grasp the overall picture of how software has changed. In contrast to the existing work, \name establishes couplings between tests and methods, \textit{directly} from how they have changed throughout the development.

Association between tests and methods can be useful in various software maintenance activities, as it can give developers hints on how their changes affect or will affect others. Consequently, there have been many studies on automatically mining this information, either dynamically or statically. Studies on test-to-code traceability links aim at setting explicit links between tests and code~\cite{White:icse:2020,Qusef:2011:icsm,Rompaey:2009:ecsmr}. Rompaey and Demeyer investigated diverse sources of information, ranging from naming convention to static and dynamic code analysis, to automatically generate traceability links between tests and code that can pinpoint which tests examine which part of code~\cite{Rompaey:2009:ecsmr}. Qusef et al. proposed SCOTCH that identifies these traceability links using dynamic slicing; SCOTCH uses dynamic slicing to locate the area affected by the last assertion in each unit test case~\cite{Qusef:2011:icsm}. 
Mohammad et al. tried to improve the quality of the traceability links by further findings the method that implements the core functionality under testing based on the changes in the object states~\cite{Mohammad:2015:scam}. Recently, White et al. combined multiple techniques that automatically mine test-to-code traceability links, allowing these techniques to complement each other~\cite{White:icse:2020}. 
Unlike all these approaches that somehow employ dynamic program analysis to generate these links between tests and code, \name is purely static. Perhaps more importantly, the traceability links focus on test intents resulting in abstract relationship between test-and-code, whereas the evolutionary couplings we use aim at capturing important dependencies.


\section{Conclusion} 
\label{sec:conclusion}

We propose \name, a static approach that mines relevant links between tests and code units without any dynamic or static code analysis but through their history/evolution. The key idea of \name is that tests and code that are relevant to each other are likely to be changed, multiple times, around the same time by developers. Thus, \name infers such relevance relationships using the past co-evolution of the software under analysis. We empirically evaluate the capability of \name in capturing such relevance relationships using 15 open-source projects. We further conducted a fault localization study to investigate the applicability and actionability of these relationships in software debugging.   
The results show that \name can establish and use such relationships and that it is capable of achieving comparable performance to an existing Information Retrieval-based fault localization technique. Further analysis reveals that \name can capture the relationships between tests and code that are different from those captured by dynamic and static code analysis and thus, evidencing that \name can complement the current approaches. 
  
 Our work forms the first attempt to establish evolutionary couplings between tests and code and  thus,  it  opens a number of interesting research directions. In particular, the exploration of additional aspects of software evolution such as actual commit time, developers and the context of the evolution could strengthen our relationships. Additionally, the formulation of hybrid techniques combining evolutionary coupling with traceability linking techniques \cite{White:icse:2020,Qusef:2011:icsm,Rompaey:2009:ecsmr} and historical evolution analysis techniques such  as refactoring miner \cite{Tsantalis:TSE:2020:RefactoringMiner2.0} could lead to much stronger results. We hope to explore these directions in the near future.   


\bibliographystyle{ACM-Reference-Format}
\bibliography{newref}

\end{document}